\documentclass[aps,twocolumn,english,superscriptaddress,citeautoscript,showkeys,preprintnumbers,amsmath,amssymb,floatfix,footinbib,prb]{revtex4-2}
\usepackage{dcolumn}
\usepackage{xcolor}
\usepackage{soul}
\usepackage{amssymb,amsmath}
\usepackage{hyperref}
\hypersetup{colorlinks=true,linkcolor=red,citecolor=blue}
\usepackage[utf8]{inputenc}
\usepackage[english]{babel}
\usepackage{txfonts}

\usepackage{graphicx} 
\usepackage{dcolumn} 
\usepackage{bm} 
\usepackage{amsmath, amssymb}
\usepackage{float}

\usepackage{chngcntr}

\addto\captionsenglish{}

\begin{document}

\title{Quantum muon diffusion and the preservation of time-reversal symmetry in the superconducting state of type-I rhenium}

\author{D. G. C. Jonas}
\email{d.jonas@warwick.ac.uk}
\affiliation{Physics Department, University of Warwick, Coventry, CV4 7AL, United Kingdom}

\author{P. K. Biswas}
\affiliation{ISIS Facility, STFC Rutherford Appleton Laboratory, Harwell Science and Innovation Campus, Oxfordshire OX11 0QX, United Kingdom}

\author{A. D. Hillier}
\affiliation{ISIS Facility, STFC Rutherford Appleton Laboratory, Harwell Science and Innovation Campus, Oxfordshire OX11 0QX, United Kingdom}

\author{D. A. Mayoh}
\affiliation{Physics Department, University of Warwick, Coventry, CV4 7AL, United Kingdom}

\author{M. R. Lees}
\email{m.r.lees@warwick.ac.uk}
\affiliation{Physics Department, University of Warwick, Coventry, CV4 7AL, United Kingdom}

\date{\today}

\begin{abstract}
Elemental rhenium exhibiting type-II superconductivity has been previously reported to break time-reversal symmetry in the superconducting state. We have investigated an arc-melted sample of rhenium exhibiting type-I superconductivity. Low temperature zero-field muon-spin relaxation measurements indicate that time-reversal symmetry is preserved in the superconducting state. Muon diffusion is observed, which is due to quantum mechanical tunneling between interstitial sites. The normal state behavior is characterized by the conduction electrons screening the muons and thermal broadening, and is typical for a metal. Energy asymmetries between muon trapping sites and the superconducting energy gap also characterize the superconducting state behavior.

\end{abstract}

\maketitle

\section{Introduction}

A transition into a conventional superconducting state is represented by the spontaneous symmetry breaking of the global $U(1)$ gauge symmetry. If any other symmetries, such as time-reversal symmetry (TRS) are broken, then the superconductor can be characterized as unconventional~\cite{Ghosh2020TRSB}. Zero-field muon-spin relaxation (ZF-$\mu$SR) measurements can detect TRS breaking because the muons interact with spontaneously generated magnetic fields which can have magnitudes dependent upon the size of the order parameter. Hence TRS breaking can be observed via an increase in the muon relaxation rate on cooling through the transition temperature, $T_{\text{c}}$. TRS breaking has been directly observed using $\mu$SR in LaNi(C, Ga)$_2$~\cite{Hillier2009LaNiC2, Hillier2012LaNiGa2}, the filled skutterudites ${\text{PrOs}}_{4}${\text{Sb}}$_{12}$~\cite{Aoki2003PrOs4Sb12} and ${\text{PrPt}}_{4}${\text{Ge}}$_{12}$~\cite{Maisuradze2010PrPt4Ge12}, and the caged compounds $({\text{Lu, Y, Sc)}}_{5}${\text{Rh}}$_{6}${\text{Sn}}$_{18}$~\cite{Bhattacharyya2015Lu5Rh6Sn18, Bhattacharyya2015Y5Rh6Sn18, Bhattacharyya2018Sc5Rh6Sn18}. However, there are conflicting results for ${\text{UPt}}_{3}$~\cite{Luke1993UPt3, DalmasdeReotier1995UPt3, Schemm2014UPt3}, the layered perovskite ${\text{Sr}}_{2}{\text{RuO}}_{4}$~\cite{Luke1998Sr2RuO4, Luke2000Sr2RuO4, Xia2006Sr2RuO4, Kashiwaya2019Sr2RuO4, Grinenko2019Sr2RuO4}, and the recently discovered ${\text{Zr}}_{3}{\text{Ir}}$~\cite{Sajilesh2019Zr3Ir, Shang2019Zr3Ir}. In single crystals of $4{\text{Hb-TaS}}_{2}$ the observation of TRS breaking is thought to be due to chiral superconductivity~\cite{Ribak20204Hb-TaS2}.

In noncentrosymmetric (NCS) superconductors, the absence of inversion symmetry in the crystal structure allows an antisymmetric spin-orbit interaction, and can lead to pairing states with a mixture of spin-singlet and spin-triplet components~\cite{Bauer2012NCSC, Smidman2017Superconductivity}. Multigap superconductivity can also be observed, for example in the TRS breaking alloy ${\text{La}_{7}}{\text{Ni}}_{3}$~\cite{Arushi2021La7Ni3}. Other NCS binary lanthanides in this family, which include ${\text{La}_{7}}{\text{(Ir, Rh, Pd)}}_{3}$, also break TRS, but show a conventional, fully-gapped $s$-wave pairing from transverse-field muon-spin rotation and heat capacity measurements~\cite{Barker2015La7Ir3, Li2018La7Ir3, Singh2018La7Rh3, Mayoh2021La7Pd3}.

Several rhenium-based superconductors, which form in the NCS $\alpha{\text{-Mn}}$ structure (space group $I\bar{4}3m$), including ${\text{Re}}_{24}{\text{Ti}}_{5}$, ${\text{Re}}_{6}{\text{(Zr, Hf, Ti)}}$, and ${\text{Re}}_{0.82}{\text{Nb}}_{0.18}$, also show the behavior of ${\text{La}_{7}}{\text{(Ir, Rh, Pd)}}_{3}$~\cite{Lue2013Re24Ti5, Shang2018Re24Ti5, Singh2014Re6Zr, Matano2016Re6Zr-Re27Zr5-Re24Zr5, Khan2016Re6Zr, Mayoh2017Re6Zr, Pang2018Re6Zr, Singh2016Re6Hf, Chen2016Re6Hf, Singh2017Re6Hf, Singh2018Re6Ti, Shang2018Re-Re0.82Nb0.12, Shang2021ReandReT}. Recent ZF-$\mu$SR measurements indicated that elemental rhenium, which is centrosymmetric (space group $P6_{3}/mmc$), also breaks TRS~\cite{Shang2018Re-Re0.82Nb0.12}. This calls into question the role the NCS structure plays in the observed behavior and alternative explanations for the TRS breaking in Re-based materials are actively being sought. For example,  electronic structure calculations show a spin-triplet component of the superconducting ground state of rhenium is allowed if spin-orbit coupling and orbital degrees of freedom are considered~\cite{Csire2020Re}.

However, there are additional observations that complicate this discussion. A linear dependence between the nuclear magnetic moment and the internal magnetic fields generated in the superconducting state for rhenium-based superconductors has been proposed~\cite{Shang2018Re-Re0.82Nb0.12}, which includes ${\text{Re}}_{3}{\text{W}}$~\cite{Biswas2012Re3W} and recently reported ${\text{Re}}_{3}{\text{B}}$ and ${\text{Re}}_{7}{\text{B}}_{3}$~\cite{Sharma2021Re7B3andRe3B, Shang2021Re7B3andRe3B}. But despite having a nuclear moment of approximately 3 nuclear magnetons, TRS is preserved in ${\text{Re}}_{3}{\text{Ta}}$~\cite{Barker2018Re3Ta}, and so does not follow this trend. Furthermore, rhenium can exhibit both type-I and type-II superconductivity. Samples that have been melted or annealed to remove internal strain are type-I, with a transition temperature $T_{\text{c}}^{\text{I}} = 1.7$~K~\cite{Hulm1957Re}. Powders on the other hand are usually reported as type-II with higher values of $T_{\text{c}}$~\cite{Shang2018Re-Re0.82Nb0.12}. Applying shear strain can produce $T_{\text{c}}$ values as high as 3.4~K~\cite{Mito2016Re}. This sample-dependent behavior motivated the work presented in this paper, to determine whether type-I rhenium breaks TRS.

When analyzing low-temperature $\mu$SR data it is often assumed that the implanted muons are static. In this case, the periodic potential barrier of the crystal traps the muons in fixed sites defined by the spatial symmetry. But muon diffusion is still possible if the transitions between sites are due to quantum mechanical tunneling~\cite{Karlsson1996QuantumDiffusion, Storchak1998QuantumDiffusion}. Conduction electrons in a metal act to screen the muons, so a combination of the bare muon and screening charge must be considered when modeling the diffusion. The presence of superconductivity complicates things further, since any inelastic scattering of the screening electrons is limited due to the formation of Cooper pairs.

In contrast to results found for type-II rhenium~\cite{Shang2018Re-Re0.82Nb0.12}, ZF-$\mu$SR measurements indicate that TRS is preserved in the superconducting state in type-I rhenium. Within the temperature range studied, we observe quantum muon diffusion, where the normal-state behaviour is typical for a metal. In the superconducting state, the diffusion can be qualitatively described by the emergence of the superconducting energy gap and energy differences between muon sites. Heat capacity, resistivity and magnetometry measurements do not indicate any unconventional behavior and the superconducting parameters are in good agreement with the literature values.

\section{Experimental details}

Disks of thickness 2-3~mm were cut from high-purity (99.99\%, Goodfellow Cambridge Ltd.) polycrystalline rhenium rods of diameter 6.4~mm. To maximally remove the internal strain, each disk was melted, flipped and remelted in an argon-arc furnace. To determine the type of superconductivity of the resulting Re buttons, magnetization ($M$) data were taken on an Oxford Instruments vibrating sample magnetometer (VSM), in applied fields up to 5~mT. Other normal state and superconducting properties were determined using a Quantum Design Physical Property Measurement System. Using the two-tau relaxation method, heat capacity ($C$) data were taken using a $^{3}{\text{He}}$ insert at temperatures from 0.5 to 3~K in fields up to 25~mT. Resistivity ($\rho$) data were taken using the four-probe alternating current method in ZF at temperatures from 1.8 to 300~K.

ZF-$\mu$SR measurements were used to investigate the nature of the superconducting state in type-I rhenium, and in particular to see if TRS is preserved. Data were taken at temperatures from 92~mK to 20~K using the MuSR spectrometer at the ISIS Pulsed Neutron and Muon Source and are available from ISIS~\cite{Jonas2020ISISData}. The Re buttons were stuck onto a 30~mm diameter silver sample holder using GE varnish. Silver foil was wrapped around the sample holder to provide a uniform thermal environment. Stray fields down to 10~$\muup{\text{T}}$ were cancelled in the sample space. A $^{3}$He-$^{4}$He dilution refrigerator was utilised to take data for temperatures $T \leq 4$~K. Positrons emitted from muon decay in the sample were detected in 64 detectors surrounding the sample space in a circular array~\cite{Lee1999Muon}. Half of the detectors are labelled forwards and the other half backwards, where the direction is defined relative to the initial muon spin. The muon asymmetry was determined from $A(t) = [N_{F}(t) - \alpha N_{B}(t)]/[N_{F}(t) + \alpha N_{B}(t)],$ where $N_{F,B}$ indicate the number of counts in the forward and backward detectors, and $\alpha$ accounts for detector efficiencies and sample thickness.

\section{Results and discussions}

\subsection{Normal and superconducting state properties}

\begin{figure}[t!]%
\centering
\includegraphics[width=\columnwidth]{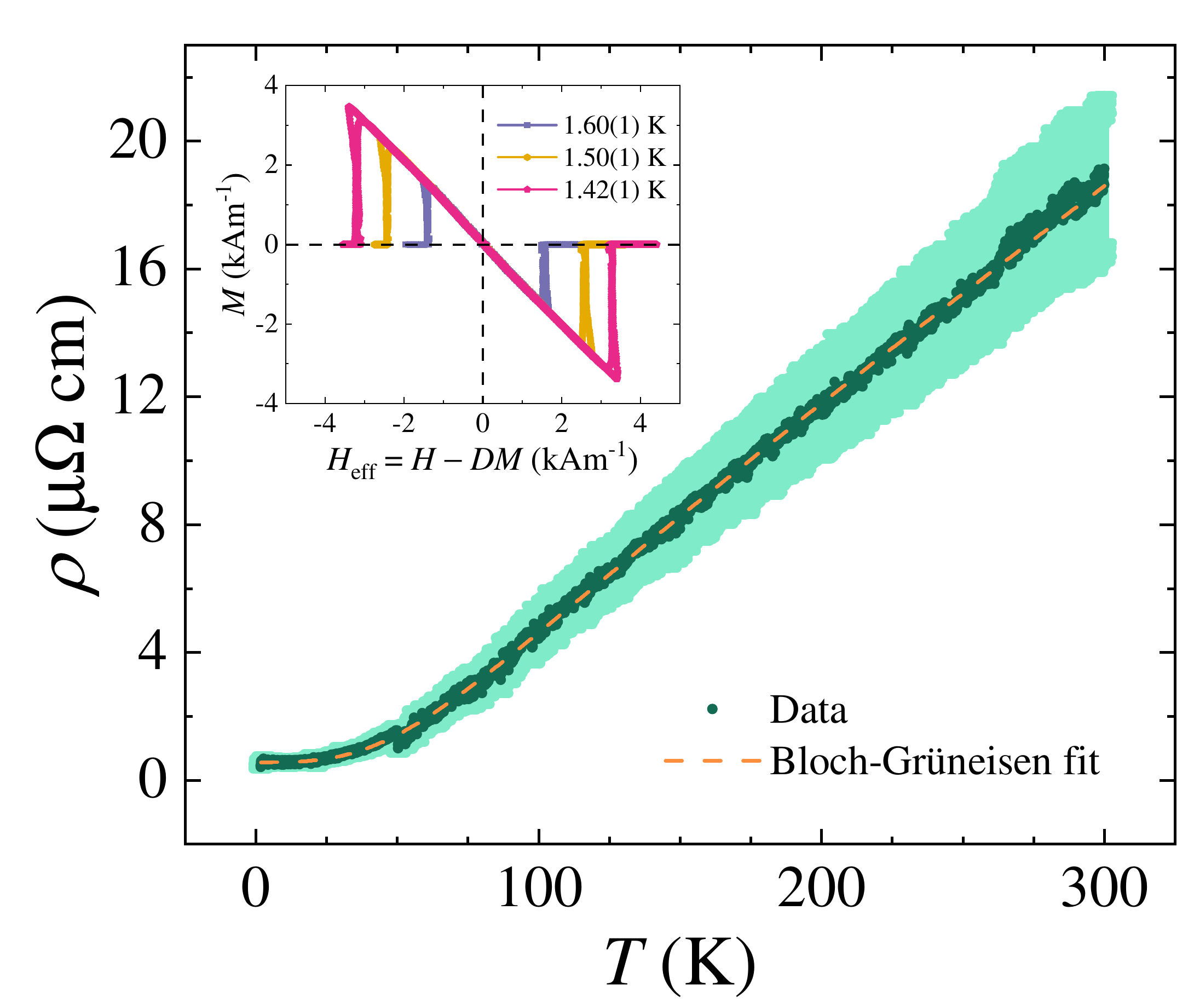}%
\caption{Normal state resistivity of rhenium as a function of temperature, $\rho(T)$. Data were fitted using the Bloch-Gr{\"u}neisen model for a transition metal~\cite{Meaden1965Resistivity}. Uncertainties in resistivity values arising from the non-uniform shape of the sample are shown by the outer band. The inset shows evidence of type-I superconductivity from magnetization versus field data, where $D = 0.29(1)$ is the demagnetization factor of the sample.}
\label{Resistivity_MvsH}%
\end{figure}

Figure~\ref{Resistivity_MvsH} shows $\rho(T)$ of rhenium. The data were fitted using $\rho(T) = \rho_{0} + \rho_{\text{BG}}(T)$, where $\rho_{0}$ is the residual resistivity due to defects and impurities, and
\begin{equation}
    \rho_{\text{BG}}(T) = 4r\left(\frac{T}{\Theta_{\text{R}}}\right)^n\int_{0}^{\Theta_{\text{R}}/T} \frac{x^n}{(e^x-1)(1-e^{-x})} \,dx
\label{Eq BG Resistivity}
\end{equation}
is the Bloch-Gr{\"{u}}neisen resistivity. The parameter $r$ is a material-dependent constant, and $\Theta_{\text{R}}$ is a characteristic temperature comparable to the Debye temperature. For transition metals where the dominant source of resistivity is from interband $s$-$d$ transitions, a value of $n = 3$ is the most appropriate~\cite{Meaden1965Resistivity}. Using this value, the fit parameters $r = 11.2(1.4)$~$\muup\Omega$~cm, $\Theta_{\text{R}} = 352(8)$~K and $\rho_{0} = 0.56(8)$~$\muup\Omega$~cm were obtained. The value of $\Theta_{\text{R}} = 352(8)$~K lies in between the Debye temperatures of rhenium at low temperatures (416~K~\cite{Smith1970Re}) and room temperature (280~K~\cite{White1957Re}). The residual resistivity ratio obtained is $\rho_{300 {\text{K}}}/\rho_{0} = 33(5)$, which is similar to values of 25-41 found for other polycrystalline samples of Re~\cite{White1957Re}. Typical magnetization data taken using a VSM are shown in the inset of Fig.~\ref{Resistivity_MvsH}. The applied magnetic field strength, $H$, was corrected using a demagnetization factor of $D = 0.29(1)$ to account for the shape of the sample using $H_{\text{eff}} = H - DM$. Type-I superconductivity is observed at each temperature studied. Curves at all temperatures are almost completely reversible, apart from a small amount of hysteresis around the critical field.

\begin{figure}[t!]%
\centering
\includegraphics[width=\columnwidth]{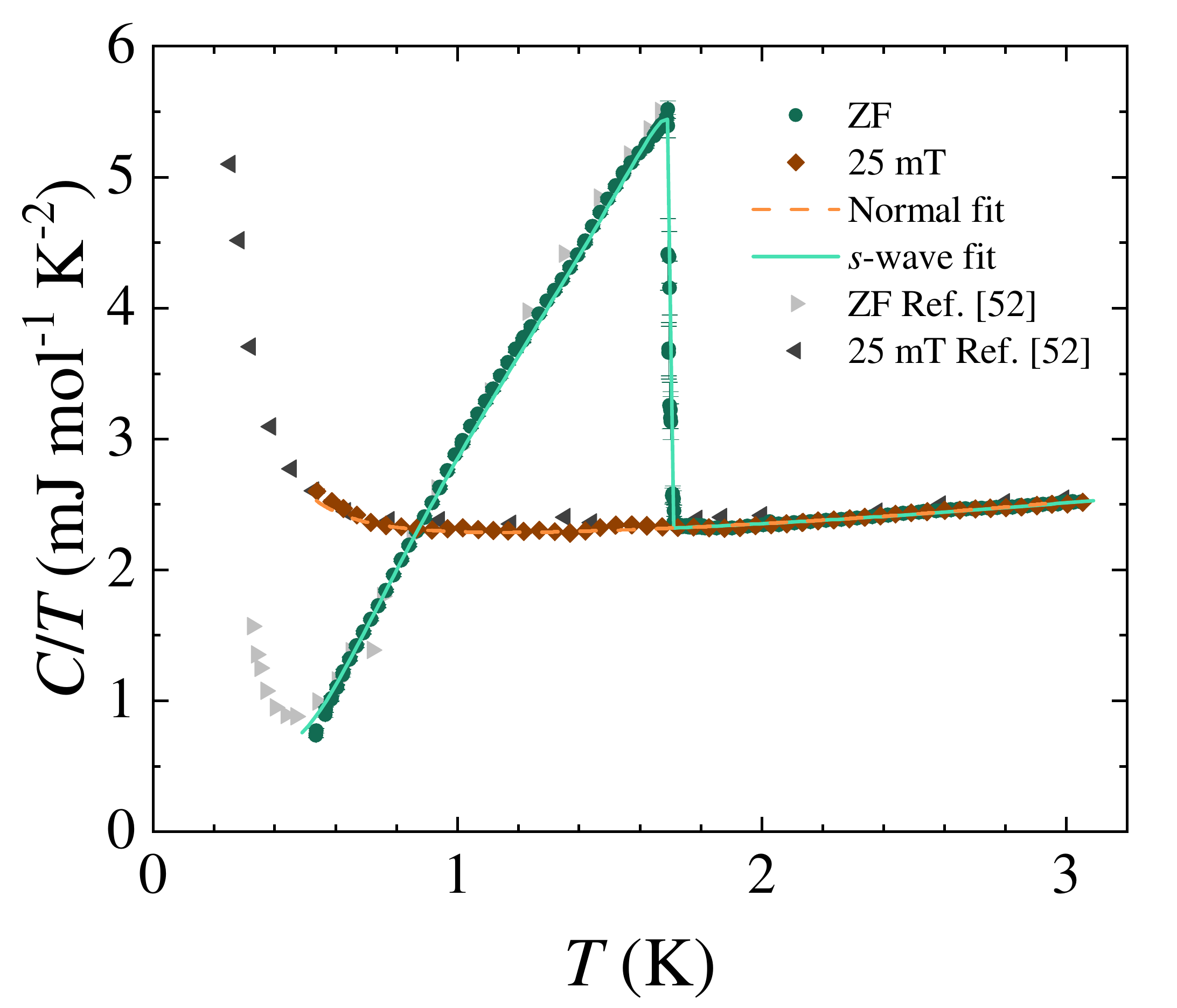}%
\caption{Heat capacity, $C$ divided by temperature, $T$ versus $T$ of rhenium. Normal state data were fit by including electronic, phonon, and Schottky anomaly contributions. Zero-field (ZF) superconducting state data were also fit using a conventional, single-gap $s$-wave model. The jump in heat capacity occurs at $T_{\text{c}}^{\text{I}} = 1.70(1)$~K, which is the reported transition temperature of type-I rhenium. Data obtained by Smith and Keesom are also shown~\cite{Smith1970Re}.}
\label{Heat Capacity}%
\end{figure}
Figure~\ref{Heat Capacity} shows $C(T)/T$ of rhenium which is in good agreement with previous work~\cite{Smith1970Re}. An applied field of 25~mT was enough to drive the sample into the normal state. The data were fitted to $C/T = \gamma_{\text{n}} + \beta T^2 + A_{\text{S}}/T^3$, where $\gamma_{\text{n}}$ is the Sommerfeld coefficient, $\beta$ is the lowest-order phonon coefficient, and $A_{\text{S}}$ determines the hyperfine Schottky-anomaly contribution. The resulting fit parameters $\gamma_{\text{n}} = 2.22(3)$~mJ~mol$^{-1}$~K$^{-2}$, $\beta = 0.032(1)$~mJ~mol$^{-1}$~K$^{-4}$ and $A_{\text{S}} = 0.047(2)$~mJ~K~mol$^{-1}$ are in good agreement with the literature values~\cite{Smith1970Re}, as expected. A standard $s$-wave model was used to fit the ZF data in the superconducting state~\cite{Bouqet2001MgB2}. The entropy $S$ was calculated from
\begin{equation}
    \frac{S}{\gamma_{\text{n}}T_{\text{c}}^{\text{I}}} = -\frac{6}{\pi^2}\frac{\Delta_{0}}{k_{\text{B}}T_{\text{c}}^{\text{I}}}\int_{0}^{\infty} [f\ln f + (1-f)\ln(1-f)] dy,
\end{equation}
where $f = [1 + {\text{exp}}(E/k_{\text{B}}T)]^{-1}$ is the Fermi-Dirac distribution and the quasiparticle energy is given by $E = \Delta_{0}\sqrt{y^2 + \delta(T)^2}$, where $y$ is the contribution from the normal-state electrons and $\Delta_{0}$ is the superconducting energy gap at $T = 0$~K. The temperature dependence of the energy gap was modeled by the expression~\cite{Carrington2003MgB2} $\delta(T) = \tanh(1.82[1.018(T_{\text{c}}^{\text{I}}/T -1)]^{0.51})$. The heat capacity was then calculated from
\begin{equation}
    \frac{C - \beta T^3 - A_{\text{S}}/T^2}{\gamma_{\text{n}}T_{\text{c}}^{\text{I}}} = T \frac{d(S/\gamma_{\text{n}}T_{\text{c}}^{\text{I}})}{dT}.
\end{equation}

\noindent Fit parameters of $A_{\text{S}} = 0.031(2)$~mJ~K~mol$^{-1}$ and $\Delta_{0}/k_{\text{B}}T_{\text{c}}^{\text{I}} = 1.74(1)$ were obtained. The Schottky constant is smaller than expected, and the gap is larger than the value of $\Delta_{0}/k_{\text{B}}T_{\text{c}}^{\text{I}} = 1.715$ previously obtained~\cite{Smith1970Re}. Nevertheless, the data are well described by a conventional $s$-wave model. Having carefully verified the type-I superconductivity in these samples, $\mu$SR spectroscopy was used to investigate the reported TRS breaking in rhenium. 

\subsection{Zero-field muon-spin relaxation}

\begin{figure}[t!]%
\centering
\includegraphics[width=\columnwidth]{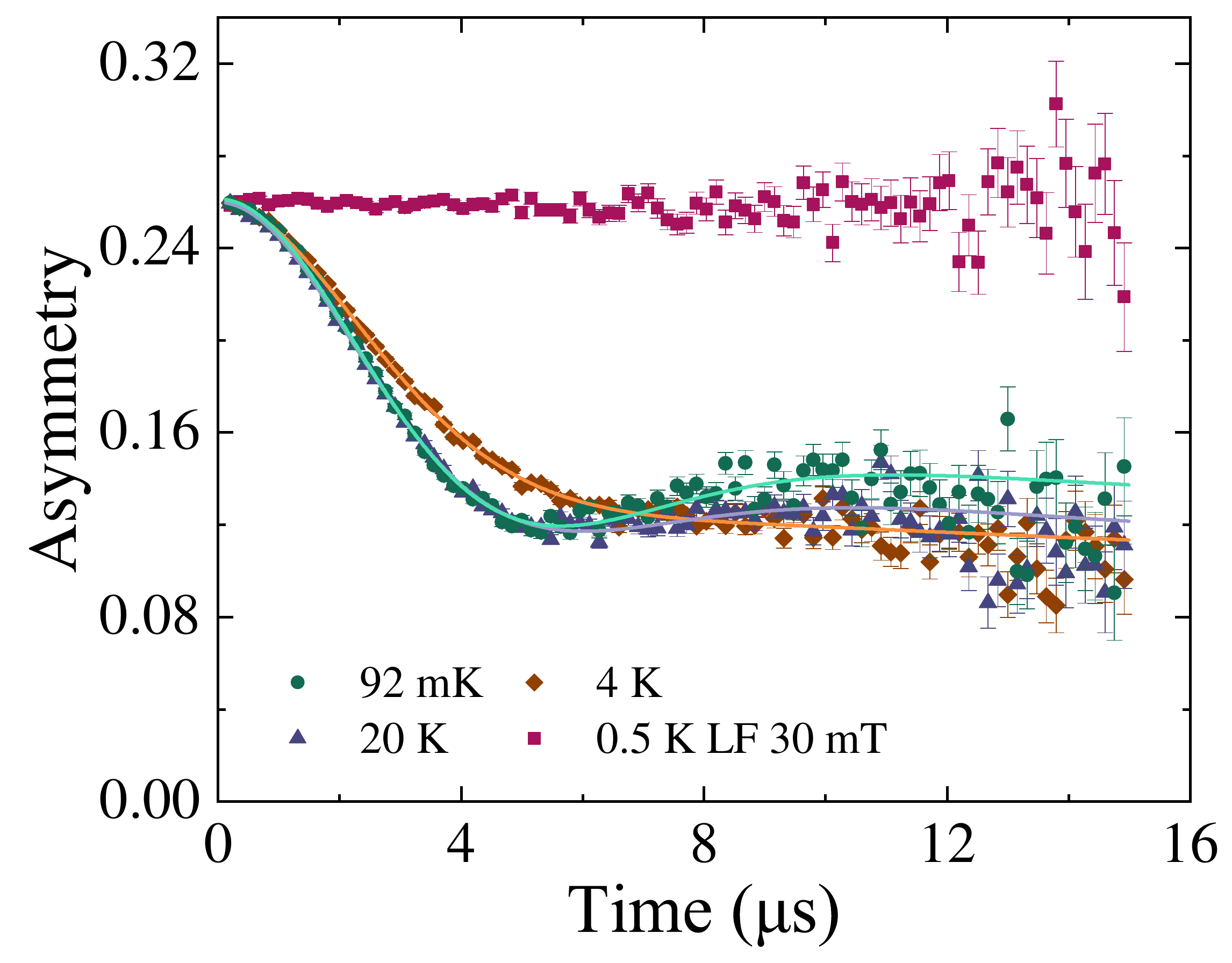}%
\caption{Zero-field muon-spin relaxation spectra for type-I rhenium above and below the superconducting transition temperature $T_{\text{c}}^{\text{I}} = 1.7$~K. Fits to each temperature curve were found using Eq.~\ref{Eq Asymmetry Diffusion} and include static relaxation effects and muon diffusion~\cite{KuboToyabe1967Magnetic, Hayano1979ZF-LF-SpinRelaxation}. Longitudinal field (LF) data in 30~mT are also shown.}%
\label{Asymmetry Raw}%
\end{figure}

Figure~\ref{Asymmetry Raw} shows ZF-$\mu$SR asymmetry time-spectra taken in the normal and superconducting states of type-I rhenium. In the presence of randomly oriented nuclear dipole moments, the asymmetry can be modeled by the Kubo-Toyabe relaxation function, given by~\cite{KuboToyabe1967Magnetic}
\begin{equation}
    A_{\text{KT}}(t) = A_0\left[\frac{1}{3}+\frac{2}{3}(1-\sigma^2t^2)e^{-\sigma^2t^2/2}\right],
\label{Eq Asymmetry Kubo-Toyabe}
\end{equation}
where $\sigma$ is the relaxation rate and $A_{0}$ is the initial asymmetry. However, by examining the 92~mK and 20~K data, a second downturn in asymmetry can be seen at later times (outside the time window of 0-8~$\muup{\text{s}}$ shown in a previous study on type-II Re~\cite{Shang2018Re-Re0.82Nb0.12}), which is indicative of muon diffusion. The data at these two temperatures are in good agreement up to $\sim$~6~$\muup{\text{s}}$, but the difference in asymmetry at later times indicates that the diffusion is temperature dependent. ZF-$\mu$SR time-spectra at all temperatures were fit within the strong collision model, with the asymmetry, $A_{\text{diff}}(t)$, given by~\cite{Hayano1979ZF-LF-SpinRelaxation}
\begin{equation}
\begin{split}
    A_{\text{diff}}(t) = A_{\text{BG}} &+ A_{\text{KT}}(t)e^{-\nu t} \\
    &+ \nu\int_{0}^{t}A_{\text{KT}}(\tau)e^{-\nu\tau}A_{\text{diff}}(t-\tau)d\tau, \\
\end{split}
\label{Eq Asymmetry Diffusion}
\end{equation}

\noindent where $A_{\text{BG}}$ is the background asymmetry from the silver sample holder, and $\nu$ is the muon hopping rate. Here, the diffusion is a Markovian process, where the muons hop between intersitial sites in the sample and relax in the same way as the static case in between hops. We report values of $A_{\text{BG}} = 0.110(2)$, $A_0 = 0.150(2)$, and $\sigma = 0.325(2)$~$\muup\text{s}^{-1}$, where the latter is in agreement with the value of $\sigma = 0.326$~$\muup\text{s}^{-1}$ found at $T = 0$~K for type-II Re~\cite{Shang2018Re-Re0.82Nb0.12}. Further information about these parameters can be found in the Supplemental Material~\cite{SuppRe}. The 4~K data in Fig.~\ref{Asymmetry Raw} show how $\nu$ can dictate the entire behavior of the asymmetry. A longitudinal field of 30~mT was large enough to remove all spin relaxation from the signal. This indicates that the internal fields are quasi-static with respect to the muon lifetime, as expected for nuclear moments.

\begin{figure}[t!]%
\centering
\includegraphics[width=\columnwidth]{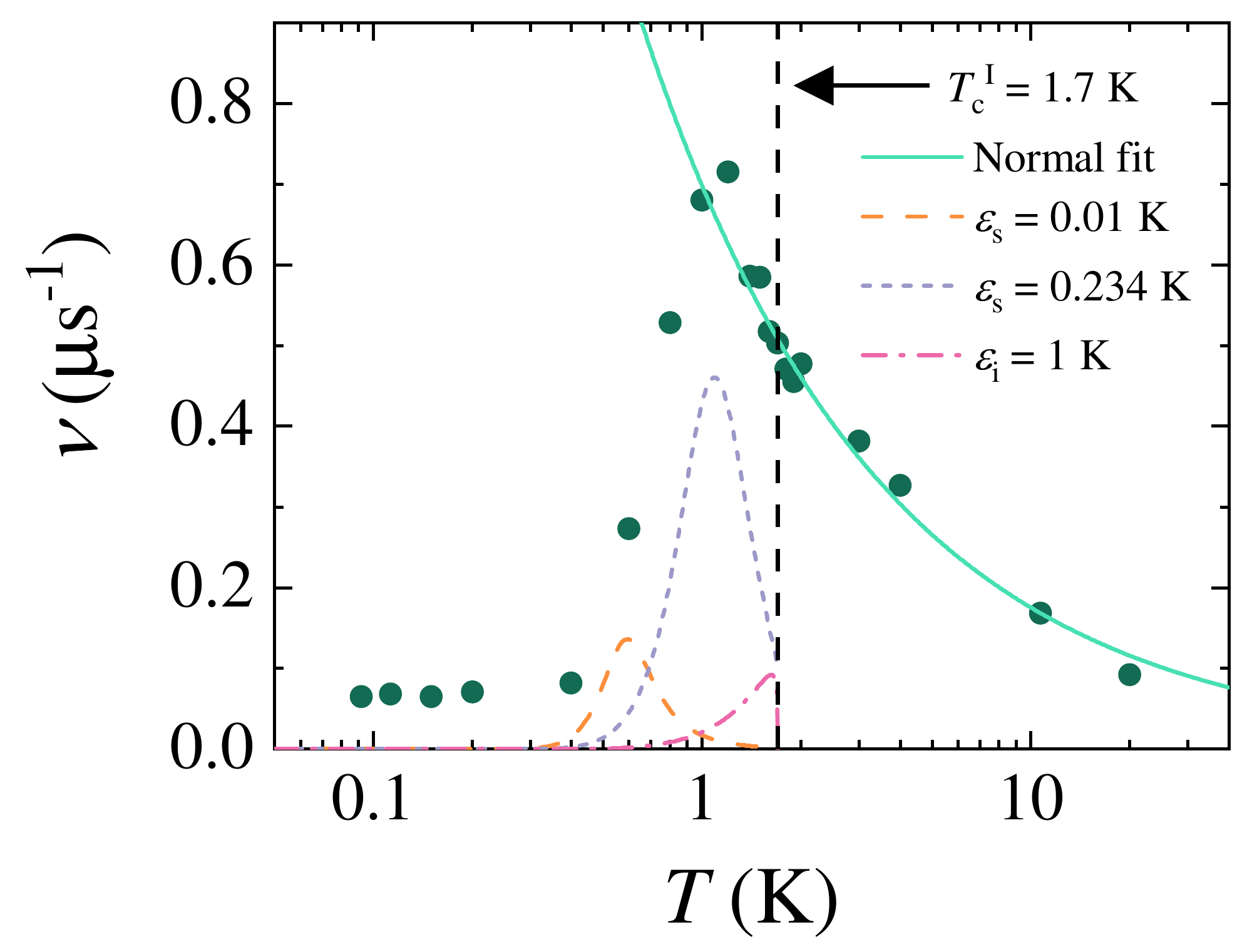}%
\caption{Muon hopping rate as a function of temperature, $\nu(T)$ where $A_{\text{BG}} = 0.11$ and $\sigma = 0.325$~$\muup\text{s}^{-1}$ were fixed in the fitting. Error bars are within the size of the data point markers. Data were fit in the normal state using $\nu(T) = a_{\text{NS}} [\Gamma(K)/\Gamma(K + 1/2)] T^{2K-1}$, where $a_{\text{NS}} = 0.20(1)$~$\muup{\text{s}}^{-1} \text{K}^{1-2K}$, and $K = 0.20(2)$. The three dashed curves below the superconducting state data were simulated using expressions for small ($\varepsilon_{\text{s}}$) and intermediate ($\varepsilon_{\text{i}}$) energy asymmetries between muon sites~\cite{Regelmann1994QuantumDiffusion}, where the energies are normalized by $k_{\text{B}}$.}
\label{Initial Asymmetry Hopping Rate}%
\end{figure}

Figure~\ref{Initial Asymmetry Hopping Rate} shows $\nu(T)$ for a fixed $\sigma = 0.325$~$\muup\text{s}^{-1}$ and $A_{\text{BG}} = 0.110$, since $A_{\text{BG}}$ is not expected to vary in the temperature range studied. The temperature axis is displayed on a logarithmic scale to show the values of $\nu$ across the full temperature range. Data were fit using~\cite{Regelmann1994QuantumDiffusion}
\begin{equation}
    \nu(T) = a_{\text{NS}} \left[\frac{\Gamma(K)}{\Gamma\left(K + \frac{1}{2}\right)}\right] T^{2K-1}
\end{equation}

\noindent in the normal state, where $a_{\text{NS}} = 0.20(1)$~$\muup{\text{s}}^{-1} \text{K}^{1-2K}$ is a constant related to the square of the tunneling matrix element and the cut-off frequency, and $\Gamma$ is the gamma function. This temperature dependence was derived by Kondo for metals, and takes into account the screening charge and thermal broadening from the surrounding electrons~\cite{Kondo1984MuonDiffusion, Kondo1986MuonDiffusion}. The dimensionless parameter $K$ is a measure of the muon-electron interaction. The obtained value $K = 0.20(2)$ is in agreement with the value of $K = 0.224(4)$ found for copper in the temperature range 1~K to 10~K~\cite{Luke1991Cu}. The reported value for aluminium is $K = 0.15$ at low temperatures, although by including a component from temperature-dependent trapping the value should be modified to $K = 0.32$~\cite{Storchak1998QuantumDiffusion}. A more comprehensive study of $\nu(T)$ across a wider temperature range would provide more insight into the behaviour of the muon in the normal state of rhenium.

The picture in the superconducting state is a complete contrast to the normal state. Since Cooper pairs cannot inelastically scatter between sites, it is more difficult for the screening charge to follow the muon, which leads to a renormalization of $K$ dependent upon the energy gap, $\Delta(T)$. Furthermore, an energy asymmetry between muon sites due to crystal defects, $\varepsilon$, means that the initial and final states of the hop are not in coherence. Subsequently, if a muon initially becomes trapped close enough to a defect, they remain trapped. The converse argument is also true due to the lack of inelastic scattering.

For a small energy asymmetry ($\varepsilon_{\text{s}}$) expressed in units of temperature, and neglecting phonon effects, the hopping rate can be determined from~\cite{Regelmann1994QuantumDiffusion}
\begin{equation}
\begin{split}
    \nu_{\text{s}}(\varepsilon_{\text{s}}, T) &= \frac{a_{\text{NS}}}{T} \cosh\left({\frac{\varepsilon_{\text{s}}}{2T}}\right) \left|\frac{\Gamma(K_{T} + \frac{i \varepsilon_{\text{s}}}{2 \pi T})}{\Gamma(K_{T})}\right|^2 \\
    &\times \left[\frac{\pi T}{2e^\gamma\Delta(T)}\right]^{2K_{T}} \left[\frac{2 e^{\gamma}\Delta(T)}{\pi}\right]^{2K} \frac{\Gamma(K_{T})}{\Gamma\left(K_{T} + \frac{1}{2}\right)}\\
    & \times e^{-2K \times \delta C(T)}[1+w(K, \varepsilon_{\text{s}}, T)], \\
\end{split}
\end{equation}

\noindent where $K_{T} = 2 K/[1+e^{\Delta(T)/T}]$ is the renormalized muon-electron parameter, $\gamma = 0.5772$ is Euler's constant, and $\Delta(T)$ is the energy gap. For the analysis relevant to this work, the energy gap used was $\Delta(T) = 1.764 T_{\text{c}}^{\text{I}} \delta(T)$, which is the conventional single $s$-wave energy gap normalized by $k_{\text{B}}$. Both $\delta C(T)$ and $w(K, \varepsilon_{\text{s}}, T)$ were set to zero, however, doing so does not affect the qualitative behavior of $\nu_{\text{s}}(\varepsilon_{\text{s}}, T)$, which is the focus of this discussion. For intermediate energy asymmetries ($\varepsilon_{\text{i}}$), the hopping rate from perturbation analysis is modified to~\cite{Regelmann1994QuantumDiffusion}
\begin{equation}
    \nu_{\text{i}}(\varepsilon_{\text{i}}, T) = a_{\text{NS}} \left(\frac{2 e^{\gamma}\Delta(T)}{\pi}\right)^{2K} \sqrt{\frac{\pi^{4} K_{T}^{2} T}{\varepsilon_{\text{i}}^3 \left(1+\frac{\varepsilon_{\text{i}}}{2 \Delta(T)}\right)}}.
\end{equation}

It is important to note that physical systems do not have one value of $\varepsilon$, because a bulk sample will always contain defects. Hence a range of $\varepsilon$ values would be present and each muon will have likely experienced different values at different points in its lifetime. Nevertheless, the qualitative behavior can be compared. For the fit parameters $a_{\text{NS}} = 0.20(1)$~$\muup{\text{s}}^{-1} \text{K}^{1-2K}$ and $K = 0.20(2)$, simulated curves of $0.2\nu_{\text{s}}(0.234$~${\text{K}}, T)$ and $0.0005\nu_{\text{s}}(0.01$~${\text{K}}, T)$ are shown in Fig.~\ref{Initial Asymmetry Hopping Rate}, which correspond to 20\% and 0.05\% of the muons experiencing the specified $\varepsilon_{\text{s}}$ throughout the entire motion. In both cases, the hopping rate increases to a maximum with decreasing $T$, before decreasing to zero, which agrees with the qualitative behavior of the data down to 0.4~K. The peaks in $\nu_{\text{s}}(\varepsilon_{\text{s}}, T)$ originate from the quasiparticle excitations in the system. Just below $T_{\text{c}}^{\text{I}}$, the fraction of Cooper pairs compared to quasiparticle excitations is small, which means that the $\nu_{\text{s}}$ is not suppressed compared to the normal state. However, as the temperature is lowered, the decrease in normal electrons means that the muons have less electrons available that can inelastically scatter. Since $\varepsilon_{\text{s}}$ is too small to cause any Cooper pair breaking processes, the muon decouples from the quasiparticle excitations (due to the trapping potential from the defects that cause $\varepsilon_{\text{s}}$), the diffusion is suppressed, and $\nu_{\text{s}}$ decreases.

A simulated curve of $0.3\nu_{\text{i}}(1$~${\text{K}}, T)$ is also shown in Fig.~\ref{Initial Asymmetry Hopping Rate}, corresponding to 30\% of the muons experiencing $\varepsilon_{\text{i}} = 1$~K. This demonstrates the situation where the muon decouples from the quasiparticle excitations and the hopping becomes limited as the sample enters the superconducting state.

Between 92~mK and 0.4~K, $\nu$ is temperature independent. Within the model used to simulate $\nu_{\text{s}}(\varepsilon_{\text{s}}, T)$, a smaller percentage of muons is needed to produce the same $\nu_{\text{s}}$ maximum for a smaller $\varepsilon_{\text{s}}$, which also shifts the peak to a lower temperature. This can be seen by the prefactors present in the examples shown. But applying this argument down to the lowest temperatures does not have any physical meaning. Alternatively, if the muons are split into a fraction that is trapped and cannot hop, and another fraction that initially stops far enough away from defects that it cannot get trapped, a constant $\nu$ indicates that the same fraction of muons are not trapped.

\section{Summary and conclusions}

Quantum muon diffusion is the key to understanding ZF-$\mu$SR measurements on rhenium that exhibits type-I superconductivity, where TRS is preserved in the superconducting state. Whilst $\nu(T)$ can be described quantitatively in the normal state, where metallic behavior is observed, in the superconducting state muon diffusion is a complicated process in which several competing mechanisms contribute to the observed behavior. Qualitatively however, $\nu(T)$ data are well described if $\varepsilon$ and $\Delta(T)$ are included in the discussion. These results require a reconsideration of TRS breaking in elemental rhenium, and demonstrate the importance of quantum muon diffusion when analyzing muon spectroscopy data at low temperatures. It would be interesting to investigate Re samples with varying degrees of strain (defects) to determine whether muon diffusion is significant in determining the form of ZF-$\mu$SR spectra for type-II Re.

Data will be made available via Warwick Research Archive Portal~\cite{WRAPRe}.

\begin{acknowledgments}
The authors would like to thank T. E. Orton and P. Ruddy for valuable technical support. This work is supported by the UKRI and STFC through the provision of beam time at the ISIS Neutron and Muon Source, UK. This work is funded by the EPSRC, UK, through Grant No. EP/R513374/1.
\end{acknowledgments}

\bibliography{David_Jonas_SC_References} 

\newpage


\counterwithin{figure}{section}
\renewcommand{\thefigure}{S\arabic{figure}}
\renewcommand{\thetable}{S\arabic{table}}
\renewcommand{\theequation}{S\arabic{equation}}

\section{Supplemental Material}

When analysing the muon spectroscopy data, the correlations between pairs of fit parameters were considered. In Eqs. \ref{Eq Asymmetry Kubo-Toyabe} and \ref{Eq Asymmetry Diffusion}, both $A_{\text{BG}}$ and $A_0$ should be temperature independent. However, letting both parameters vary in the fitting lead to incorrect fits, because changes in $A_{\text{BG}}$ are compensated by changes in $A_0$, which produce lower reduced chi-squared values. 
\begin{figure}[!h]
\centering
\includegraphics[width=\columnwidth]{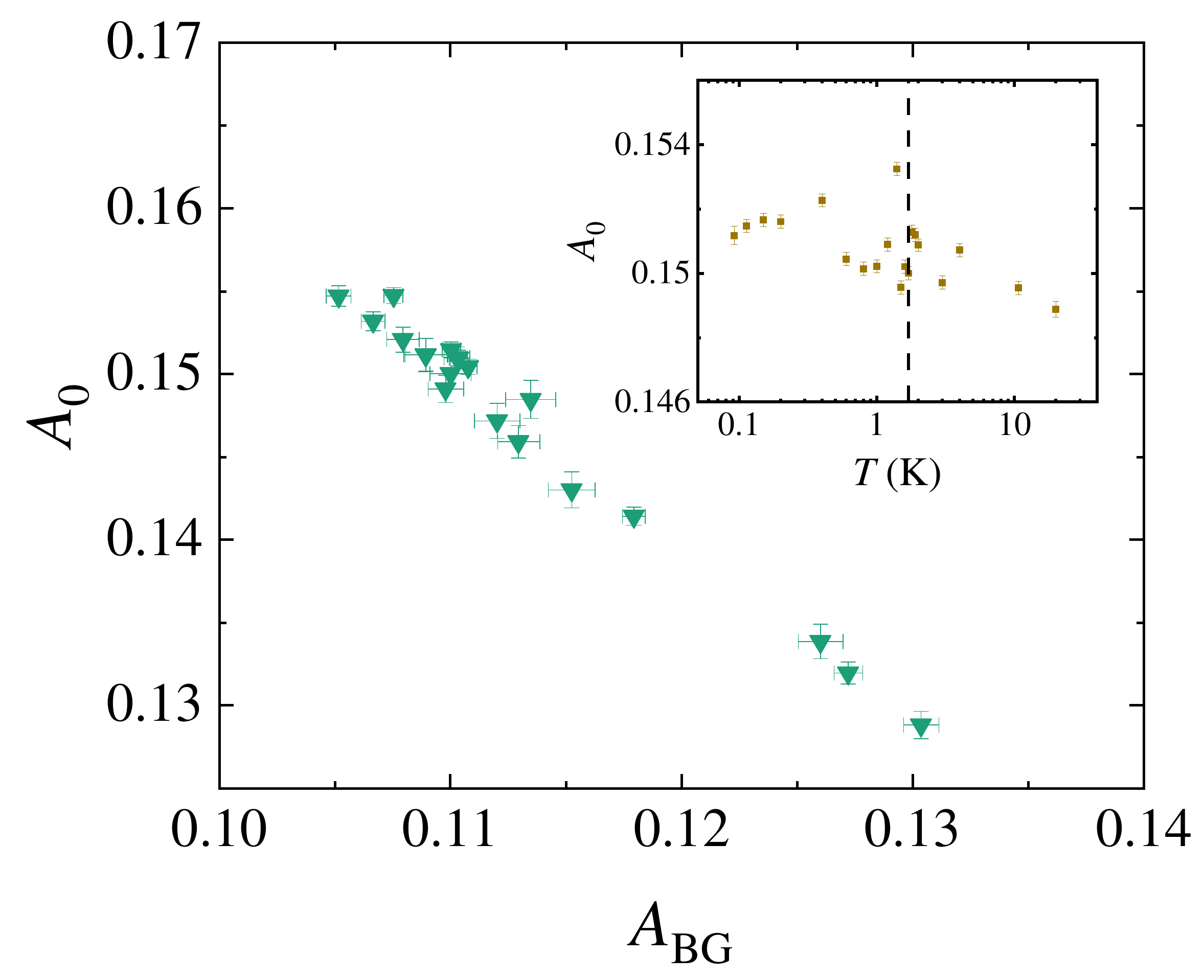}
\caption{Initial asymmetry, $A_0$ against the background asymmetry, $A_{\text{BG}}$. The Pearson correlation coefficient for $A_0$ and $A_{\text{BG}}$ is $-0.99$, which indicates that $A_0$ and $A_{\text{BG}}$ are strongly correlated. By fixing $A_{\text{BG}} = 0.11$ and $\sigma = 0.325$~$\muup\text{s}^{-1}$ in the fitting, which is the same fitting regime used for Fig. 4 in the main text, the correlation can be removed, as shown in the inset of $A_0$ versus temperature, $T$. The dashed line in the inset is the transition temperature of type-I rhenium, $T_{\text{c}}^{\text{I}} = 1.7$~K.}
\label{IABA Figure}
\end{figure}

This is illustrated in Fig.~\ref{IABA Figure}. A linear relationship between $A_{\text{BG}}$ and $A_0$ can be seen, with a corresponding Pearson correlation coefficient of $-0.99$. Hence the background asymmetry was fixed to $A_{\text{BG}} = 0.11$, which is the value at 92~mK, when making conclusions about the data. The large value of $A_{\text{BG}} = 0.11$ is due to the presence of gaps between the solid buttons of rhenium that formed the sample. The inset in Fig.~\ref{IABA Figure} shows that when fixing $A_{\text{BG}} = 0.11$ and the relaxation rate to $\sigma = 0.325$~$\muup\text{s}^{-1}$, $A_0$ is approximately temperature independent, as expected.

\begin{figure}[!b]
\centering
\includegraphics[width=\columnwidth]{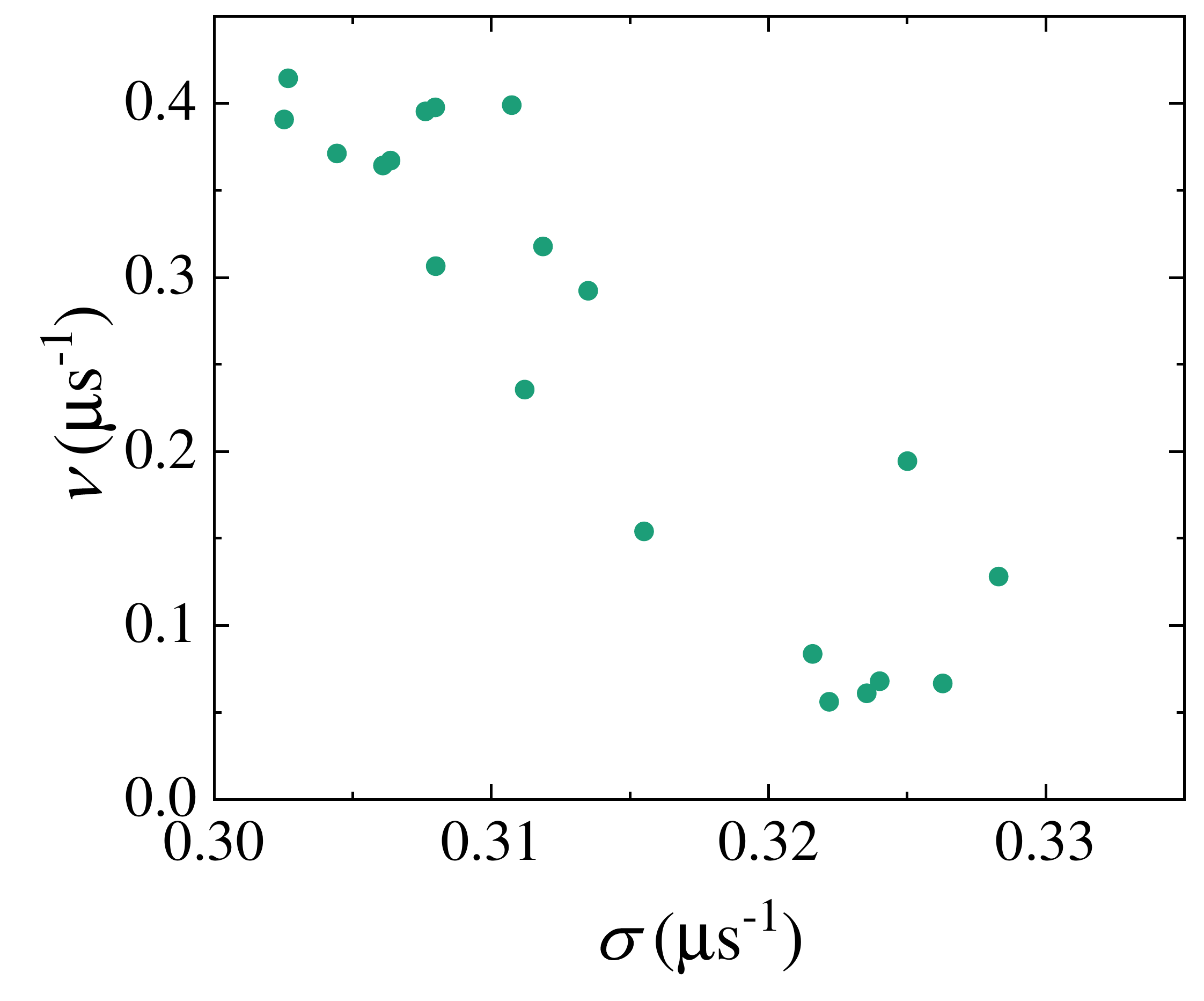}
\caption{Hopping rate, $\nu$ versus relaxation rate, $\sigma$. A Spearman correlation coefficient of -0.84 indicates that $\nu$ and $\sigma$ are highly correlated.}
\label{RR HR Correlations Figure}
\end{figure}

\begin{figure}[!t]
\centering
\includegraphics[width=\columnwidth]{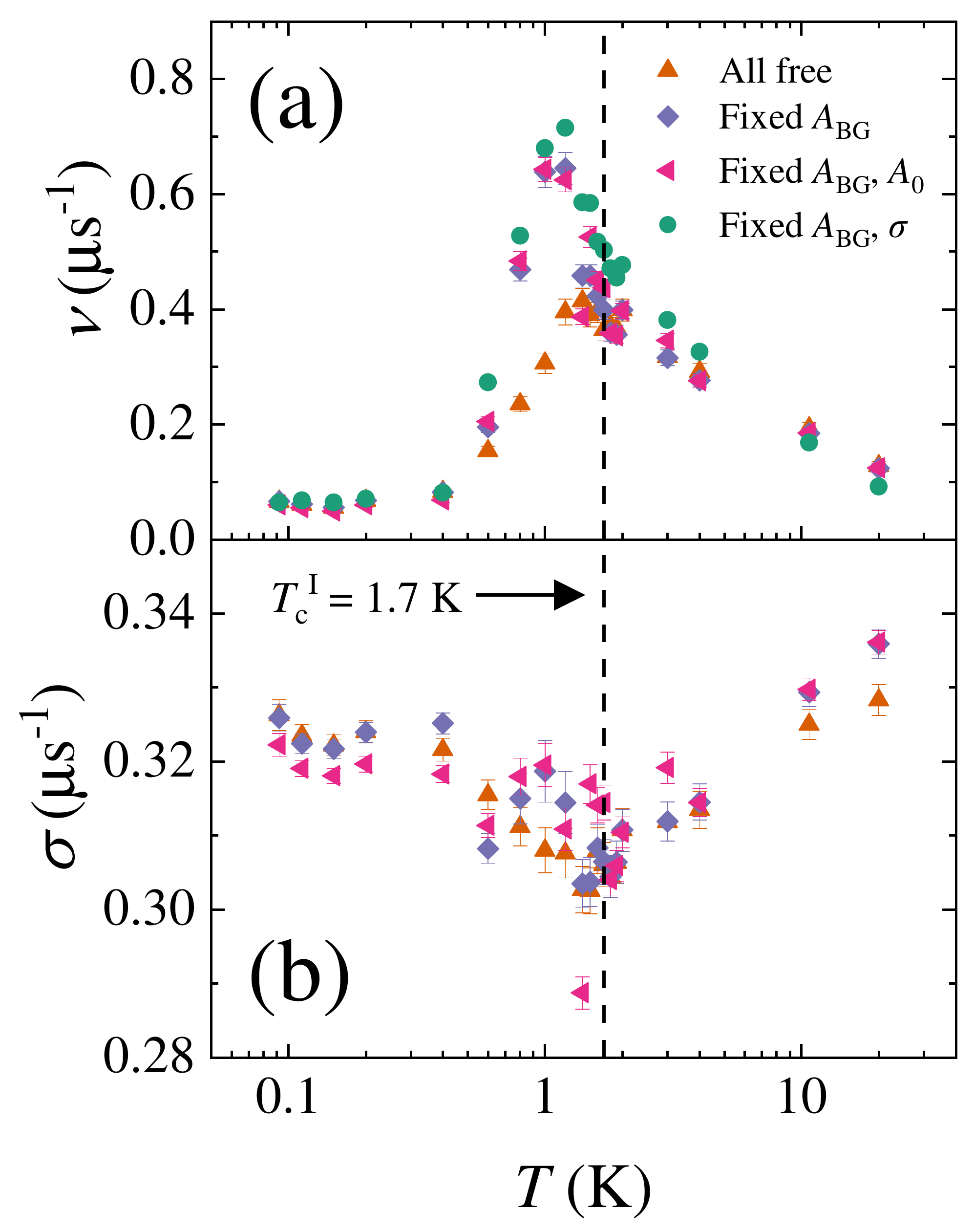}
\caption{(a) Muon hopping rate as a function of temperature, $\nu(T)$ found using Eqs.~\ref{Eq Asymmetry Diffusion} and \ref{Eq Asymmetry Kubo-Toyabe}~\cite{Hayano1979ZF-LF-SpinRelaxation}. Fit parameters that were fixed in some cases are $A_{\text{BG}} = 0.11$, $A_{0} = 0.15$, and $\sigma = 0.325$~$\muup\text{s}^{-1}$. (b) Corresponding relaxation rate values, $\sigma(T)$. Fixing the background and initial asymmetries results in slight differences in $\sigma$ compared with allowing all the parameters in the fit to vary. The transition temperature, $T_{\text{c}}^{\text{I}}$, of type-I rhenium is shown.}
\label{HR RR Figure}
\end{figure}

Figure~\ref{RR HR Correlations Figure} shows that the hopping rate, $\nu$ and $\sigma$ are also highly correlated. The Spearman correlation coefficient for the two parameters is $-0.84$. This is expected because both parameters have similar contributions inside and outside the exponentials in Eqs.~\ref{Eq Asymmetry Kubo-Toyabe} and \ref{Eq Asymmetry Diffusion}. To gain a better understanding of how to interpret the data, $\nu$ and $\sigma$ were studied in different fitting regimes, with data shown in Fig.~\ref{HR RR Figure}.

Figure~\ref{HR RR Figure}(a) shows $\nu(T)$ for a range of fitting regimes. The values that were fixed in some cases are $A_{\text{BG}} = 0.110$, $A_0 = 0.150$, and $\sigma = 0.325$~$\muup\text{s}^{-1}$. Apart from the case where all parameters were allowed to vary, the qualitative behavior of $\nu(T)$ does not depend on the parameters that were fixed. This implies that the temperature dependence is intrinsic to this sample of rhenium.

Figure~\ref{HR RR Figure}(b) shows $\sigma(T)$ in the same fitting regimes as Figure~\ref{HR RR Figure}(a). Above $T_{\text{c}}^{\text{I}}$ there is a clear increase with temperature in all the data sets. However, in the normal state the nuclear dipole moments and hence $\sigma$ should be temperature independent. Below the superconducting transition temperature of rhenium, $T_{\text{c}}^{\text{I}} = 1.7$~K, an increase in $\sigma$ can be seen, which if viewed in isolation may suggest the presence of time-reversal symmetry breaking. But since the values of $\sigma$ coincide at the lowest and highest temperatures, we conclude that the apparent temperature dependence shown in Fig.~\ref{HR RR Figure}(b) is due to the correlation of $\sigma$ and $\nu$, and $\sigma$ should be fixed in the fitting. Hence we conclude that time-reversal symmetry is preserved in the superconducting state of type-I rhenium, and report a value of $\sigma = 0.325(2)$~$\muup\text{s}^{-1}$.

\end{document}